\documentclass[12pt]{amsart}

\usepackage{amsmath,amssymb, amsthm, amsfonts}
\usepackage{enumerate}
\usepackage{xcolor}
\usepackage{tikz}
\usepackage{graphics}
\usepackage{calc}
\usepackage{marginnote}
\usepackage{caption}
\usepackage{subcaption}
\usepackage{mathabx,epsfig}

\usepackage[utf8]{inputenc}

\usepackage{hyperref}
\hypersetup{ colorlinks = true,linkcolor = {blue},citecolor = {red} }

\theoremstyle{definition}
\newtheorem{remark}{Remark}

\newtheorem{example}{Example}

\newcommand{\RR}{\mathbb{R}}

\newcommand{\OO}{\mathcal O}
\newcommand{\g}{\mathfrak g}

\newcommand{\be}{\begin{equation}}
\newcommand{\ee}{\end{equation}}

\newcommand{\VS}[1]{{ \framebox{VS:}~#1~\framebox{:VS}}}
\newcommand{\OS}[1]{{ \framebox{OS:}~#1~\framebox{:OS}}}

\begin{document}
\title[]{Odd Wilson surfaces}

\author{Olga Chekeres}
\address{ITMP, Lomonosov Moscow State University, Leninskie Gory, GSP-1, Moscow, 119991, Russian Federation}
\email{olya.chekeres@gmail.com}
\author{Vladimir Salnikov}
\address{LaSIE UMR 7356 CNRS / La Rochelle University,
Avenue Michel Crépeau, 
F-17042 La Rochelle Cedex 1 - France
}
\email{vladimir.salnikov@univ-lr.fr}

\begin{abstract}
 Previously, Wilson surface observables were interpreted as a class of Poisson sigma models. We profit from this construction to define and study the super version of Wilson surfaces. We provide some `proof of concept' examples to illustrate modifications resulting from appearance of odd degrees of freedom in the target. \\

\end{abstract}

\maketitle

\thispagestyle{empty}
\section{Introduction}

Wilson lines, loops and surfaces are important examples of non-local observables in gauge theory. This paper is the first in a series, where we intend to explore natural generalizations of the Wilson surface observable construction of \cite{OC18, ACM15}. The first non-trivial near-at-hand step is the the supersymmetric setting. More precisely, we consider adding extra degrees of freedom to the target, which can eventually be defined on a super space or super manifold, mimicking thus some supersymmetrization procedure.

The Wilson surface (WS) observables are originally defined on integral coadjoint orbits of compact Lie groups. 
One of the remarkable results of \cite{ACM15}, before addressing the quantization questions, is the idea that up to an appropriate redefinition of fields the resulting theory can be viewed as a Poisson sigma model (PSM -- \cite{PSM}). The respective (linear) Poisson structure comes from the Lie algebra structure, or more precisely from its dual, which is used to define the target.

In more detail, this paper is essentially oriented to analysis of Lie super groups/algebras instead of smooth (i.e. purely even) ones used in the original construction. While the general guideline seems rather straightforward: `make sure that the super analogues of all the ingredients are well-defined, and assemble them to observables' as before, there are some details worth being spelled out. For example, the standard difficulty lies in the integration of differential forms; and the explicit computations are  more intricate for physically relevant cases. Here the interpretation in terms of Poisson structures provides an important geometric insight.  

The paper is organized as follows:
First, we recall the results about Wilson surface observables and the standard construction of the Poisson sigma model from Wilson surfaces (section \ref{sec:PSM}). While the result there is known, we still do it with a certain pedagogical effort, since this is precisely the construction needed for further generalizations in this paper and its sequels. 
 In section \ref{sec:SuWi} we add supersymmetry into the play, namely consider super Lie groups instead of smooth ones. In the original non-super case the relation to the Poisson sigma model was an observation that allowed to compute partition functions of Wilson surface theory for all compact Lie groups. Here it becomes an even more powerful tool to avoid some ill-defined (or technically subtle) supergeometric constructions. Examples of section \ref{sec:examples} illustrate possible qualitative outcomes of this procedure, in particular we pay attention to how `non-trivial' is the odd part of the resulting theory.
The main message of the paper is to explain the subtleties related to super geometry in the context, which we state in the main text, but for self-consistency of it, some minimal working knowledge about supermanifolds and Lie supergroups/algebras is recalled in the Appendix \ref{sec:superman}.

\section{From Wilson loops to PSM}  \label{sec:PSM}

\subsection{Wilson line and surface definitions} \label{subsec:wilson-poisson} 

\subsubsection*{General gauge theory set up} 
Let $G$ be a compact connected Lie group, $\g=Lie(G)$ its Lie algebra, $<\cdot ,\cdot >$ an invariant product on $\g$. $G$ plays the role of a structure group for a gauge theory on a smooth orientable manifold $\Sigma$, a short hand for it would be ``$G$ is the gauge group''.
The relevant geometric structure is a principal $G$-bundle $P\to \Sigma$. The gauge connection $A$ is given by a $\g$-valued 1-form on $P$.

\subsubsection*{Wilson line}  
Historically, the \emph{Wilson line observable} is defined  
by a curve 
$\Gamma: [0,1]\to \Sigma$ and some representation $R$ of the gauge group $G$:
$$
W_\Gamma^R=Tr_R Pexp\left(\int_\Gamma A_R\right),
$$
where $Pexp$ stands for the path ordered exponential, which is a way to compute the holonomy of the gauge connection,
and taking the trace $Tr_R$ in the representation $R$ guarantees gauge invariance.

This construction was revisited by Alekseev, Faddeev and Shatashvili who suggested in \cite{AFS} a new description based on path integral quantization of symplectic phase space: 
\begin{equation} \label{W-gamma}
W_\Gamma^R=\int \mathcal{D}g e^{S[A,g,\lambda]},
\end{equation}
where the integral is taken over all  the maps from the curve $\Gamma$ to the gauge group $G$: $g: \Gamma \to G$, conveniently avoiding the need of path ordering, and  the action functional is
\be
\label{AFS}
S[A,g,\lambda]=\int_\Gamma <\,\lambda, g^{-1}dg+g^{-1}Ag>=\int_\Gamma <\, b,gdg^{-1}+A>.
\ee
This is a 1-dimensional sigma model with a coadjoint orbit\footnote{For more details on geometry of $\g^*$ and $\OO_\lambda$ see Appendix \ref{sec:orbits}.  } $\OO_\lambda$ as a target.  The fields in the construction are as follows: $\lambda \in \g^* $ is a constant orbit representative, the field $g: \Gamma \to G$ takes values in the gauge group, and $b:\Gamma \to \g^*$ is an auxiliary field with the property $b(t)=g(t)\lambda g(t)^{-1}$, $t$ being a parameter on the curve $\Gamma$.

\subsubsection*{Toward Wilson surface}  
Further developping the theory, Diakonov and Petrov noticed (\cite{DP}) that via Stokes theorem, one can rewrite the functional \eqref{AFS} as an integral over the surface $\Sigma$ enclosed by the curve $\Gamma$.
In \cite{ACM15} the following formula was obtained for Diakonov--Petrov action: 
$$
S[A,g,\lambda]=\int_\Sigma d \, <\, \lambda,g^{-1}dg+g^{-1}Ag>=\int_\Sigma d\, < \, b,gdg^{-1}+A>.
$$
After a simple calculation the resulting action functional becomes:
\begin{eqnarray}
\label{DP}
S[A,g,\lambda]=\int_\Sigma <b,F_A-\frac{1}{2}[dgg^{-1}+A,dgg^{-1}+A]> = \nonumber
\\ 
=\int_\Sigma <b,F_A - (d_Agg^{-1})^2>,    
\end{eqnarray} 
where $F_A=dA+\frac{1}{2}[A,A]$ and $d_Ag= dg +Ag$.

In \cite{ACM15} formula \eqref{DP} was interpreted as an equivariant extension of the Kirillov--Kostant--Souriau symplectic form, allowing thus to construct an observable 
on any closed 2d orientable surface. The key distinction here is that this sigma model is not (necessarily) defined as a boundary sigma model. We will consider this as a starting point for constructing the analog of the theory in the setting of superalgebras.  

\begin{remark}
    In \eqref{DP} the functional still depends on a ``parameter'' $\lambda$ that labels the orbits, it is hidden in the $b$ field. The functional is in principal  defined for any ``value'' of $\lambda$. 
    Although for the present paper it is not particularly important, as a side remark, let us note that in \cite{ACM15} only special values of $\lambda$ were considered, corresponding to integral orbits -- this is related to the Kirillov's orbit method and to geometric quantization.
\end{remark}

\subsection{Wilson surface as a Poisson sigma model} \label{sec:sigma-models} $\;$ \\

As mentioned above, one of the messages of \cite{ACM15} is that there is a natural way to view the  functional of Wilson surface observable as a Poisson sigma model. While in the original work it was not necessarily explored in details, it will be crucial to have in mind for the present paper and the sequels, we thus sketch the construction here.

Let $\Sigma$ be a 2-dimensional orientable surface, possibly with several boundary components. $G$, $A$, $b$, $F_A$, etc as above.  An auxiliary horizontal (in the usual sense of equivarant cohomology) 
$1$-form field $\alpha \in \Omega^1_{hor}(P, \g)^G$ is introduced to modify the action \eqref{DP} in the following way:
\begin{eqnarray}
S^\pi(b,g,A,\alpha) =\int_\Sigma < b, F_A + (d_Agg^{-1} +\alpha)^2- (d_Agg^{-1})^2>=
\nonumber \\ \label{WSP}    
=\int_\Sigma <b, d(A+\alpha)+(A+\alpha)^2>.
\end{eqnarray}

The two actions are not equal, but equivalent, in the sense that path integrals are equal, that is when the auxiliary field $\alpha$ is integrated out, the theory reduces to \eqref{DP}:
\be \label{Wilson-PSM} \nonumber
\int \, \mathcal{D}g \, \mathcal{D}\alpha \, e^{S^\pi(b,g,A,\alpha)} =\int \, \mathcal{D}g \, e^{S(b,g,A)}.
\ee
Several interesting observations can be made about \eqref{WSP}. First, $A+\alpha$ is a new connection on $P$. The 2-form $F_{A+\alpha}=d(A+\alpha)+(A+\alpha)^2$  is its curvature. 
Second, one now sees rather clearly that it can be viewed as a particular case of the Poisson sigma model. 

As a reminder,  the Poisson sigma model (\cite{PSM}) is defined on vector bundle morphisms between $T \Sigma$ for some $\Sigma$ -- oriented two dimensional manifold, and $T^*M$, where $(M, \pi)$ is a Poisson manifold. These morphisms can be encoded in two types of fields: scalar $X \colon \Sigma \to M$, and 1-form $A\in \Omega^1(\Sigma, X^* T^* M)$.
The functional reads: 
\be \label{PSM-functional}
S^\pi = \int_{\Sigma} \Big(AdX + \frac{1}{2} X^*\pi(A)\Big)
= \int_{\Sigma} \Big(A_i dX^i + \frac{1}{2} \pi^{ij}A_i A_j\Big).
\ee

Equations of motion are obtained as extrema of \eqref{PSM-functional}:
\begin{equation} \label{PSM-eom} 
dX^i + \pi^{ij} A_j = 0, \qquad  dA_i + \pi^{jk}_{,i}A_j A_k = 0
\end{equation}
And one checks that \eqref{PSM-functional} is invariant under the following gauge transformations:
\begin{equation} \label{PSM-gt}
\delta_{\varepsilon}X^i = \pi^{ij} \varepsilon_j, \qquad  \delta_{\varepsilon}A_i = d\varepsilon_i + \pi^{jk}_{,i}A_j 
\varepsilon_k 
\end{equation}

A concise way to view the Wilson surface observable as a PSM consists of two steps of simplifications of the latter. First, the considered Poisson structure is linear, it is given by the standard construction\footnote{See again Appendix \ref{sec:orbits}.} on the dual to a Lie algebra $\mathfrak{g}^*$; i.e. $\pi^{ij} = C^{ij}_k x^k$, the coefficients being the structure constants of the Lie algebra.
Second, the target in the Wilson surface setting is a fixed coadjoint orbit $\OO_\lambda$, so this Poisson structure is restricted to it. 
This means that that the resulting $\Pi_{red}$ is non-degenerate and is the inverse of a symplectic form on $\OO_\lambda$. A simple computation shows that the form of equations of motion, symmetries, etc, remains the same in this restricted setting; and we will drop the subscript ``red'' in what follows.

\begin{remark}
    The second step of simplification is not strictly necessary: one can just consider Wilson surfaces with an enlarged target as linear Poisson sigma models. We will explore this idea in the context of interacting Wilson surfaces in \cite{OCVS-interaction}. 
\end{remark}

\begin{remark}
While the general perception of linear Poisson structures is rather as mathematically exotic creatures, the discussion above shows that they have interesting physical meaning, despite their simplicity. 
Note that, the constructed $\pi_{red}^{-1}$ in the generic situation is not necessarily in Darboux form. But if the coordinate change to bring it to the canonical form is known (like for $SU(2)$ and $SU(3)$ -- \cite{AFS}), this gives some explicit information for the corresponding PSM.
    \end{remark}

\begin{remark} 
  Continuing the previous remark, the linear Poisson sigma model is equivalent, as expected, to a 2d $BF$-theory: 
$$
S=\int_\Sigma <b, F_{A+\alpha} >\\
$$
Here, the field $b\in\g^*$ is constrained to take values in the orbit. Wilson surface can be interpreted as an independent 2d topological theory with 1d Hilbert space. 
The equations of motion, from variation with respect  to $b$ and
from variation with respect to the connection $A+\alpha$ (up to a sign):
are in immediate analogy with the regression to linear Poisson $\pi^{ij} = C^{ij}_kX^k$ in \eqref{PSM-eom}, as well as gauge transformation reproduce \eqref{PSM-gt}.
\end{remark}

The following example will be important for the super version of the Wilson surface observable, so we give it in detail. 

\begin{example}  \label{example:SU}
    Let $G=SU(2)$, the algebra $\mathfrak{su}(2)\cong \RR^3$. The Poisson bracket on $\RR^3$ is $\{x^i,x^j\}=\epsilon^{ijk}x^k$, $\epsilon$ being the Levi-Civita symbol. The symplectic leaves (orbits) are $\OO_\lambda \cong S^2$, each sphere has radius corresponding to a Casimir function $R^2=x^ix^i$ (summation over repeated indices is implied). Orbits are labeled not only by elements $\lambda\in \mathfrak{h}^*$, but also by Casimir functions. This is expected, since Casimir functions are invariants of representations. 
The Poisson structure reads:
\begin{equation} \nonumber
    \pi = \epsilon^{ijk}x^k \frac{\partial}{\partial x^i}\frac{\partial}{\partial x^j}.
\end{equation}
The restriction of this structure to an orbit does not make it vanish, it just implies that there is a constraint on the generators: $x^ix^i=const$.
The corresponding symplectic structure is:
\be \nonumber
\omega=\frac{1}{2R^2}\epsilon^{ijk}x^idx^jdx^k.
\ee
\end{example}

\section{Supersymmetrized Wilson surface observable} \label{sec:SuWi}

 Let us now turn to the extension of the Wilson surface: adding the super degrees of freedom. As expected, this means 
 considering Lie supergroups instead of smooth (purely even) ones. In this section we provide only the upshot of the construction, paying attention to the subtleties that occur. 
We suppose that the reader is familiar with super geometry, if this is not the case, some  notions are recalled in Appendix \ref{sec:superman}.

\subsection{Towards the definition.}
The first natural attempt would be to go through all the steps of the previous section adding the word ``super'' everywhere. While theoretically doable, this procedure is rather technical: the structures on Lie groups/algebras, as well as their links to representation theory is now less straightforward, and some concepts need to be redefined. We will thus focus directly on constructing the super analog of 2-dimensional model \eqref{DP}. 

The starting point of our construction in the super case is the definition \eqref{WSP} of the Wilson surface in terms of Poisson sigma-model. 
That is, we define the functional of the \emph{supersymmetric Wilson surface} to be the super Poisson sigma model with a linear super Poisson structure on the target space. 
We consider the identification of the data of dual to a Lie superalgebra $\g^*$ with some linear super Poisson structure $\pi$, which works almost verbatim as in the smooth case, obviously up to signs in the identities.
\begin{remark}  \label{rem:ibort} 
The standard $\g^*$-like example of a super Lie--Poisson structure (\cite{ibort}) contains a subtlety:  for a Lie supergroup $G$ 
with the Lie superalgebra $\g = \g_0 \oplus \g_1$,
one considers the superalgebra $\g = \g_0^* \oplus \g_1$, i.e.  reads off the Poisson structure on ($\g_0^*, S(\g_1)$). In what follows we will mainly use some modifications of it.    
\end{remark}
  Then, the coadjoint orbits should correspond to symplectic leaves, and those, equipped with the restriction of the super Poisson structure $\pi_{red}$ will be viewed as the target of the theory.  
  This result is much more subtle in the super case: the Weinstein splitting theorem is not automatic. But for our purposes it is enough to consider one orbit, which, according to \cite{tilmann}, is super symplectic. Constructing a super Wilson surface observable as a super Poisson sigma-model on super symplectic leaves has two important advantages: it provides a great source of examples and it is based on well-defined classical super geometry. 

\begin{remark}
When we say that super PSM is well-defined, we do not mean that it is a simple construction. The original paper \cite{sPSM} only treated its physical applications (for supergravity), not going into supergeometric details. It has been revisited in \cite{superAKSZ} with a more or less by hand construction of the mapping spaces. A more conceptual approach would be to view sPSM in the context of $Q$-bundles (\cite{q-bundle}). Even the simplified (linear) case we consider, should be properly defined as a functional between multigraded (super) manifolds. By this we mean that the rigorous way to describe all the fields is to view them as mappings between manifolds graded by a monoid ($\mathbb Z_2 \times \mathbb N$), which is a straighforward regression of \cite{AKVS-z-graded}.
\end{remark}

\begin{remark} \label{rem:holonomy}
   As  mentioned above, it is in principal possible to perform the supersymmetrization procedure along the lines of section \ref{sec:PSM}, starting from Wilson loops.  
   This relates to an interesting purely mathematical question of holonomy on supermanifolds (\cite{galaev}), which we are going to explore elsewhere.  
\end{remark}

\subsection{Super WS action functional}
We define the super WS functional as a particular case of super PSM, i.e. the equation \eqref{PSM-functional} with $\pi$ being a linear super Poisson structure. For applications and concrete examples it is still important and convenient to keep track of the relation to the BF form of the model. The supersymmetric  
functional is then structurally the same as non supersymmetric:
\begin{equation} \label{superWS}
S=\int_\Sigma Tr\Big(\mathbf{bF}_{\mathbf{A+a}} \Big).\\    
\end{equation}
Now $G$ is a matrix Lie supergroup with a Lie superalgebra  $\g = \g_0\oplus\g_1$. The fields (we write them in bold letters) are now superfields. The gauge potential is a $1$-form taking values in Lie superalgebra $\g$: \\
$\mathbf{A}=A^{a\xi}_idx^i\otimes T_a\otimes e_\xi\in \Omega^1(P,\g_0\oplus\g_1)$. The field $\mathbf{b}: \Sigma \to \g_0^*\oplus\g_1^*$ is constrained to take values in the super orbit, with the property $\mathbf{b}=(g,\alpha)(X,\beta)(g,\alpha)^{-1}$, where $(g,\alpha)\in G$ and $ (X,\beta) = \lambda\in \mathfrak{h}_0^*\oplus\mathfrak{h}_1^*$  
is the orbit representative with $\mathfrak{h}_0^*\subset \g^*_0$ being the dual to the Cartan subalgebra of $\g_0$, and $\mathfrak{h}_1^* = \Pi \mathfrak{h}_0^*$ being the odd extension of $\mathfrak{h}_0^*$ ($\Pi$ stands for shifted parity); \\
$\mathbf{a} \in \Omega^1_{hor}(P, \g)^G$, as before, is an auxiliary field, given by a $G$-horizontal 
$1$-form with values in the Lie superalgebra $\g$. $Tr$ stands for invariant bilinear product on the Lie superalgebra $\g= \g_0\oplus\g_1$.

\section{Examples and interpretation} \label{sec:examples}

Let us recall the effect of adding the Wilson surface observable to a background 2-dimensional gauge theory. In  \cite{OC18} and \cite{ACM15} the modification of 2-dimensional Yang-Mills partition function due to the WS has been computed for various compact Lie groups. 

For the case $G=U(1)$ the partition function is modified by a prefactor depending on the orbit representative $\lambda$. %
For $SO(3)$ the partition function is split into 2 terms, each of them is multiplied by its own prefactor, again depending on $\lambda$. The same phenomenon occurs for other compact Lie groups: the result is given by summation over all classes of the principal G-bundles over the source and each term is multiplied by the corresponding WS phase. Not to overload the message with unnecessary details about the test theory, we will only consider the partition function of a ``pure'' supersymmetric Wilson surface observable. Namely, 
we illustrate how the mentioned prefactors may behave depending on the choice of the super group. 

As mentioned above, the definition of the functional via super Poisson structure provides a class of geometric examples -- one of them (totally artificial, section \ref{sec:twistedSU2}) will serve as a warm up exercise to observe the supersymmetrization phenomenon. 
A more systematic approach (sections \ref{sec:SUSU} and \ref{sec:OSP}) consists in extending the group or extending the space it is acting on. As in the non-super case, an important class of examples comes from quadractic Lie algebras, where, in particular, we do not have to worry about the remark \ref{rem:ibort}. For the first such example we will only study the orbit space, and for the second one we will also compute partition functions.

\subsection{``Twisted'' SU(2)} 
\label{sec:twistedSU2}
In the usual $SU(2)$ case the orbit space is very explicit: $\mathbb R^3$ is foliated by two-dimensional spheres. Then, recall that in the construction of Wilson surface the orbit parameter $\lambda$ labels these spheres and can be related to the radial (transversal) coordinate. 

Consider now the same foliation, but declare formally the radial component to be odd. This corresponds to a ``twist'' of the $\mathfrak{su}(2)$ Lie bracket, making it odd. Or, alternatively, a similar construction can be obtained from the Heisenberg group with one odd generator. 
A possible interpretation of this construction is also a change of parity in the Cartan subalgebra of $\g$.
Regardless of the origin of the foliation, we can just view it geometrically. Recall now, that in \cite{ACM15} $\lambda$ was entering the game in the exponential prefactor of the modified partition function, and for integral orbits for simply connected groups
was not producing any modification. Now, when we formally consider it to be an odd element, the exponential reduces to $(1+\lambda)$, which is always a non-trivial super phase prefactor. 

A similar construction may be carried out in a more general setting of $SU(N)$ -- in contrast with non-super case, the partition function will be modified by a non-trivial super phase.

\subsection{Extension of SU(2) by its odd algebra} 
\label{sec:SUSU}
This example is in a sense inspired by the theory of odd quadratic superalgebras (\cite{odd-quadratic}), and in particular the extension of $sl(2)$, we adapt it to the supergroup $ G = SU(2)  \ltimes \Pi\mathfrak{su}(2)$, and its Lie algebra $\g=\mathfrak{su}(2)\oplus \Pi \mathfrak{su}(2)$. Here $\Pi$ stands for shifted parity. We will compute adjoint orbits for this group with the purpose to explore their structure. In the usual (non-super) case they would have been isomorphic to coadjoint ones, for a superalgebra the relation is again more intricate, we will comment on that below.

Let us first recall the form of the operations on $G:=G_0 \ltimes \Pi\g_0$. The elements of $G$ are pairs $(g,\alpha)\in G$, where $g\in G_0$, $\alpha\in \Pi\g_0$. 
The group product is given (as for usual semi-direct products) by 
\begin{equation}  \label{su-pi-su-prod}  \nonumber
(h,\beta)\cdot (g,\alpha)=(hg, \beta +h\alpha h^{-1}). 
\end{equation}
Then the inverse $(g,\alpha)^{-1}=(g^{-1}, - g^{-1}\alpha g)$ is defined from a simple calculation: 
$$
(h,\beta)\cdot (g,\alpha)=(hg, \beta +h\alpha h^{-1})=(I, 0).
$$
The left and right actions are 
$$
L_{(g,\alpha)}(h,\beta)=(gh, \alpha + g\beta g^{-1}),
$$
$$
R_{(g,\alpha)}(h,\beta)=(hg, \beta + h\alpha h^{-1}).
$$
The Lie algebra of $G$ is $Lie(G)=\g_0\oplus \g_1$, where $\g_0=Lie(G_0)$ and $ \g_1= \Pi Lie(G_0)$.
The adjoint action is defined by
\be \nonumber
  Ad_{(g,\alpha)}=R_{(g,\alpha)^{-1}}\circ L_{(g,\alpha)} \, \, at \,\, (I, 0).
\ee
For $(X,\beta) \in Lie(G)$  it is computed as follows: 
\be   \nonumber
\begin{split}
   Ad_{(g,\alpha)}(X,\beta)&= \left.\frac{d}{dt}\Big(R_{(g,\alpha)^{-1}}\circ L_{(g,\alpha)}(e^{tX},t\beta)\Big)\right|_{t=0} \\
  &=\Big(gXg^{-1}, g\beta g^{-1}+[gXg^{-1},\alpha]\Big)\\
  &=(g,\alpha)(X,\beta)(g,\alpha)^{-1}.
\end{split}
\ee
We will now specialize these computations for particular elements, to describe the orbit space. \\
We are interested in potential correspondence between orbits and representations, that is why we consider orbits passing through elements $(X,\beta)\in \mathfrak{h}_0\oplus \Pi \mathfrak{h}_0$. \\

\textit{Adjoint orbits with an even representative} \\
For $(X,0)$, the stabilizer consists of elements $(g,\alpha)$ such that \\
$(X,0) = Ad_{(g,\alpha)}(X,0)=\Big(gXg^{-1}, [gXg^{-1},\alpha]\Big)$.

For $X\in \mathfrak{u}(1)$, the conditions $X=gXg^{-1}$ and $[X,\alpha]=0 $ are satisfied when $g\in U(1)$ is in maximal torus subgroup of $SU(2)$, and  $\alpha\in \Pi\mathfrak{h}_0= \Pi\mathfrak{u}(1)$ is in Cartan subalgebra of $\mathfrak{su}(2)$ with shifted parity. \\
The super stabilizer is: $\mathcal{H}_{(X,0)}=U(1)\ltimes \Pi\mathfrak{u}(1).$ \\
The corresponding orbit, as a manifold, is: $$\mathcal{O}_{(X,0)}\cong SU(2)\times \Pi\mathfrak{su}(2) / U(1)\times \Pi\mathfrak{u}(1)\cong S^2\times \Pi \mathbb{R}^2.$$
And for $X=0$ the stabilizer is obviously the entire group \\ $\mathcal{H}_{(0,0)}=SU(2)\ltimes \Pi\mathfrak{su}(2)$. \\

\textit{Adjoint orbits with an odd representative} \\
For $(0,\beta)$, the stabilizer consists of elements $(g,\alpha)$ such that \\ $(0,\beta) = Ad_{(g,\alpha)}(0,\beta)=\Big(0, g\beta g^{-1}\Big)$. \\
The element is $\beta \in \Pi\mathfrak{u}(1)$. The only condition is $\beta=g\beta g^{-1}$, and it is satisfied when $g\in U(1)$ is in maximal torus subgroup of $SU(2)$, and  $\alpha\in  \Pi \mathfrak{su}(2)$ is any element of $\Pi \mathfrak{su}(2)$. \\
The stabilizer is: $\mathcal{H}_{(0,\beta)}=U(1)\ltimes \Pi\mathfrak{su}(2).$ \\
As a manifold, the orbit is $$\mathcal{O}_{(0,\beta)}\cong SU(2)\times \Pi\mathfrak{su}(2) / U(1)\times \Pi\mathfrak{su}(2)\cong ``S^2 \times \text{odd vector}''.$$
\begin{remark}
Let us explain the last equality in the above line. When computing the quotients, we obtain the result up to an isomorphism of (super)manifolds. While topologically the result is a sphere (viewed as a subset of $G_0$), it clearly has only an odd part. Hence imagining $\OO_{(0, \beta)}$ as a set of points is misleading, the proper way is to consider the sheaf of functions on it, which would be equivalent to functions on $S^2$ up to an odd (constant) prefactor. 
\end{remark}

\textit{Adjoint orbits with a generic representative.} \\
For $(X,\beta)$, the stabilizer consists of elements $(g,\alpha)$ such that \\$(X,\beta) = Ad_{(g,\alpha)}(X,\beta)=\Big(gXg^{-1}, g\beta g^{-1}+[gXg^{-1},\alpha]\Big)$. \\
The element is $(X,\beta) \in \mathfrak{u}(1)\oplus \Pi\mathfrak{u}(1)$. The conditions for a stabilizer are $\beta=g\beta g^{-1}$, $X=gXg^{-1}$ and $[gXg^{-1},\alpha]=0 $. They are satisfied when $g\in U(1)$ is in maximal torus subgroup of $SU(2)$, and  $\alpha\in \Pi\mathfrak{h}= \Pi\mathfrak{u}(1)$ is in Cartan subalgebra of $\mathfrak{su}(2)$ with shifted parity. \\
The stabilizer is
$\mathcal{H}_{(X,\beta)}=U(1)\ltimes \Pi\mathfrak{u}(1).$ \\
As a manifold, the orbit is $$\mathcal{O}_{(X,\beta)}\cong SU(2)\times \Pi\mathfrak{su}(2) / U(1)\times \Pi\mathfrak{u}(1)\cong S^2\times \Pi \mathbb{R}^2.$$ 
While the computation is slightly different, one notices that the even and the generic case coincide, and are essentially different (even for dimension reasons) from the odd case. This means that the orbit space indeed defines a non-trivial singular (super)foliation. \\

\textit{The functional.} \\
Let us now come back to the construction of the WS functional. 
There is no automatic identification of adjoint and coadjoint orbits, but it may be done in two steps: first identify $\Pi \g_0$ and $\Pi (\g_0^*)$, then use a bilinear form on the whole algebra. The subtle point is that the resulting ``identification'' is an odd one, i.e. parity reversing. 
This is however enough for our illustrative purposes -- we see two types of orbits: one with mixed and another with pure parity. 
This means that such a super extension of $SU(2)$ produces two essentially different classes of super extensions of WS functionals. \\

\textit{Generalization to compact Lie groups.} \\
Note that the above computation can be performed for any compact Lie group extended by its odd algebra.\\
 For an even/generic representative, 
the stabilizer of  $(X,\beta)\in \g_0\oplus \Pi \g_0$ is $\mathcal{H}_{(X,\beta)}=H_0\ltimes \Pi\mathfrak{h}_0.$
\\
The orbit is $\mathcal{O}_{(X,\beta)}\cong G_0\times \Pi\g_0 / H_0\times \Pi\mathfrak{h}\cong G_0 / H_0 \times \Pi \mathbb{R}^{dim(\g_0)-dim(\mathfrak{h}_0)},$
where $H_0 \subset G_0$ is a maximal torus subgroup of $G_0$ and $\mathfrak{h}_0\subset \g_0$ is a Cartan subalgebra of $\g_0$.
\\
For an odd representative,
the stabilizer of  $(0,\beta)\in \g_o\oplus \Pi \g_0$ is \\ $\mathcal{H}_{(0,\beta)}=H_0\ltimes \Pi\g_0.$ \\
The orbit is, as in a similar case of $SU(2)$, a subspace of the odd subalgebra $\Pi \g_0$, which is, as a geometric figure, isomorphic to $G_0/H_0$: $$\mathcal{O}_{(0,\beta)}\cong G_0\times \Pi\g_0 / H_0\times \Pi\mathfrak{g}_0\cong ``G_0 / H_0 \times odd \, \, vector''.$$
And once again, after identifications of corresponding dual algebras, this produces non-trivial WS functionals. 

\begin{remark}
Interpreting the above construction in terms of Poisson geometry, we expect a non-trivial relation to tangent Poisson structures, see also \cite{superAKSZ}.

\end{remark}

\subsection{Coadjoint orbits of OSp(1$|$2)} 
\label{sec:OSP}

$OSp(1|2)$ is a classical Lie supergroup, its algebra admits a bilinear invariant non-degenerate form. Since this form is even, we can identify $\mathfrak{g}\cong \g^*$. 

Coadjoint orbits of $OSp(1|2)$ are super discs: $\mathcal{D}^{(2|2)}\cong OSp(1|2)/ U(1)$ (\cite{GR95}). They admit an even super symplectic form  and 
 are superextensions of a one-parametric family of elliptic $Sp(2,\RR)\cong SU(1,1)$-orbits: $\mathcal{D}^2\cong SU(1,1)/U(1)$.
Geometric quantization of $\mathcal{D}^2$ yields highest weight representations of $SU(1,1)$. The procedure extends to $\mathcal{D}^{(2|2)}$ to yield irreducible super unitary representations of $OSp(1|2)$.

$OSp(1|2)$ is the simplest super extension of $Sp(2,\RR)\cong SU(1,1)$. $SU(1,1)$ is the maximal even subgroup of $OSp(1|2)$. 
We consider the compact real form of $OSp(1|2)$ (see \cite{Ber}), denoted by  $UOSp(1|2)$ which is a supersymmetric extension of the compact simply connected group $Sp(1)=SU(2)=Spin(3)$. 

The algebra $\mathfrak{uosp}(1|2)$ has the following structure: the even (bosonic) generators $h$, $b_+$, $b_-$ are those of $\mathfrak{su}(2)$, and the odd (fermionic) generators $f_+$, $f_-$ are the basis of the odd subspace $\Pi \mathfrak{p_0}$, where $\mathfrak{p_0}$ is the invariant compliment of the Cartan subalgebra $\mathfrak{h_0} \subset \g_0$ in the Cartan decomposition $\g_0=\mathfrak{h}_0\oplus \mathfrak{p}_0$, $h$ is the generator of $\mathfrak{h}_0$. Then $\mathfrak{uosp}(1|2)=\mathfrak{su}(2)\oplus \Pi \mathfrak{p_0}$. 
The commutation relations read:
$$[h,b_+]=b_+, \quad [h,b_-]=-b_-, \quad [b_+,b_-]=2h, \quad [h,f_+]=\frac{1}{2}f_+,$$
$$[h,f_-]=-\frac{1}{2}f_-, \quad
[b_-,f_+]=f_-,   \quad
[b_+,f_-]=f_+, \quad
[b_+,f_+]=[b_-,f_-]=0, $$
$$[f_+,f_+]=\frac{1}{2}b_+, \quad
[f_-,f_-]=-\frac{1}{2}b_-, \quad
[f_+,f_-]=-\frac{1}{2}h.$$
The explicit matrix representation of those can be found in \cite{valya}. 
And in the physics language, this is $\mathcal{N}=1$ supersymmetry.

We use the fact that for $UOSp(1|2)$, $\g\cong \g^*$ and identify adjoint and coadjoint orbits. It is enough to consider the orbit representative $(X,0)\in \mathfrak{h}=\mathfrak{h}_0\oplus 0$ which is purely even, since Cartan subalgebra of $\g_0$ does not have an odd counterpart.
The stabilizer is purely even:  
$\mathcal{H}=U(1).$ And the orbit is 
$$\mathcal{O}_{(X,0)}=SU(2)\ltimes \Pi \mathfrak{p_0}/ U(1) \cong S^2\times \Pi \mathbb{R}^2.$$
The central elements $e$ and $-e$ are purely even, since they are elements of the maximal torus subgroup: the center $Z(G)\subset \mathcal{H}$ is a subgroup of $\mathcal{H}$. (Cartan subalgebra integrates to a maximal torus subgroup). 

\begin{remark}
Note that the orbit turns out to be the same as for the even representative of $\g=\mathfrak{su}(2)\oplus \Pi \mathfrak{su}(2)$, which may give a lead for partition functions there. 
\end{remark}

\begin{remark}
Representation theory works on these examples (\cite{kostant}), i.e. we can construct super Wilson lines, illustrating remark \ref{rem:holonomy}. But more importantly we can apply the technique from \cite{OC18} to compute partition functions, which we will do below.    
\end{remark}

\subsection{Canonical quantization of WS -- example}

Let us recall the steps described in \cite{OC18} to canonically quantize the WS action functional and obtain the partition function formula for the Wilson surface theory, and see what changes in the super case.

Supersymmetric extension of $G_0$ does not change the topology of the group. Hence, the principal $G$-bundles $P\to \Sigma$ over a closed surface are also classified by the elements $\gamma \in \pi_1(G_0)$, since the fundamental groups of $G$ and $G_0$ are the same. The gauge group can be viewed as $G=\tilde{G}/\Gamma$, where $\tilde{G}$ is the universal cover of G and $\Gamma \subset Z(\tilde{G}_0)$ is a subgroup of the center of $\tilde{G_0}$. The universal cover $\tilde{G}$ can then be described as $\tilde{G}=\tilde{G_0}\ltimes \Pi \g_0$, where $\tilde{G_0}$ is the universal cover of $G_0$. 

For each element $\gamma \in \pi_1(G)$ in the fundamental group there exists a corresponding element $C_\gamma \in \Gamma \subset Z(\tilde{G})$ in the center of the covering group, since $\Gamma \cong \pi_1(G)$. Then the partition function of the Wilson surface labeled by the element $(X,\beta)$ for a particular equivalence class of principal bundles $P\to \Sigma$, defined by $\gamma \in \pi_1(G)$ is given by:
\begin{equation}\label{phase} \nonumber
Z(C_\gamma, (X, \beta)) =\frac{\chi_{(X,\beta)}(C_\gamma)}{sD_{(X, \beta)}},    
\end{equation}
where $\chi_{(X, \beta)}(C_\gamma)=sTr(C_\gamma) $ is a value of the character $\chi_{(X, \beta)}$ on the element $C_\gamma$ in the representation $R_{(X, \beta)}$, corresponding to the orbit element ${(X, \beta)}$, and $sD_{(X, \beta)}$ is the ``super dimension'' of the matrix $R_{(X, \beta)} (C_\gamma)$  
(i.e. number of nontrivial diagonal elements of the even block minus the number of diagonal elements in the odd block in its normal form). 

Note, that we can only use this procedure to compute partition functions, when there is a bijective correspondence between some class of coadjoint orbits and finite-dimensional representations of the group $G$, i.e. when the highest weight of the representation is identified with the orbit representative. In \cite{ACM15, OC18} the examples of compact connected Lie groups were treated, and the bijective correspondence between integral coadjoint orbits and finite-dimensional irreducible representations of $G$ was explored. With super groups one has to be more careful. We apply the procedure to an example, where the orbit method is known to work. For other examples one can compute partition functions as path integrals (such examples will be treated in a separate paper), without exploring representation theory, choosing wisely the orbit representative.

\textit{$UOSp(1|2)$ partition function.} 
\\
For $G=UOSp(1|2)$ the highest weight $(X,0)\in \mathfrak{h}^*$ defines an irreducible finite-dimensional representation. In this case the target space of the theory is given by the coadjoint orbit $\mathcal{O}_{(X,0)}$ passing through the point $(X,0)\in \mathfrak{g}^*$.

For $G=UOSp(1|2)$, since there exists only the trivial class of principal $G$-bundles over  a closed surface $\Sigma$, the partition function of the Wilson surface is 
\begin{equation} \nonumber
Z(e, (X, 0)) =\frac{\chi_{(X,0)}(e)}{sD_{(X, 0)}}=\frac{sTr_{(X,0)}(e)}{sD_{(X, 0)}}=\frac{sD_{(X, 0)}}{sD_{(X, 0)}}=1,    
\end{equation}
just like in non-supersymmetric case.

For $G=UOSp(1|2)/\mathbb{Z}_2=SO(3)\ltimes \Pi \mathfrak{p}_0$ there are two classes of principal $G$-bundles over  a closed surface $\Sigma$. The partition function of the Wilson surface for the trivial class is 
\begin{equation} \nonumber
Z_{triv}=Z(e, (X, 0)) =\frac{\chi_{(X,0)}(e)}{sD_{(X, 0)}}=\frac{sTr_{(X,0)}(e)}{sD_{(X, 0)}}=\frac{sD_{(X, 0)}}{sD_{(X, 0)}}=1.    
\end{equation}
The partition function of the Wilson surface for the nontrivial class is 
\begin{equation} \nonumber
Z_{nontriv}=Z(-e, (X, 0)) =\frac{\chi_{(X,0)}(-e)}{sD_{(X, 0)}}=\frac{sTr_{(X,0)}(-e)}{sD_{(X, 0)}}=\frac{-sD_{(X, 0)}}{sD_{(X, 0)}}=-1.    \end{equation}
This result agrees with the (non supersymmetric) case of \\
$G=SU(2)/\mathbb{Z}_2=SO(3)$ computed in \cite{ACM15}, in spite of the fact that the orbit is supersymmetric.

\begin{remark}
    We would like to call this phenomenon ``bosonization'', and we believe that it happens for super extensions of compact groups, when the Cartan subalgebra of the respective super
algebra has trivial super extension. The proof of this conjecture would be technical, going through the representation theory of Lie super groups -- we do not think it is necessary in this paper.  
\end{remark}

\section*{Instead of conclusion -- perspectives}

In this paper we have seen that viewing Wilson surface observables through the lens of Poisson sigma model can be very fruitful for generalizations. We have addressed the super version of it showing that one can produce non-trivial supersymmetric modifications of the WS functional by taking into account  the information about the orbits of supergroups. We have also seen that in some cases the super contribution may disappear while performing the canonical quantization procedure. 

On top of the mentioned results, this geometric understanding of Wilson surfaces offers several open directions. First, even before studying supersymmetry, very naturally one may consider several WS observables and glue them together in the topological field theory sense. This results eventually in higher dimensional generalization of those (\cite{OCVS-interaction}). Second, we have studied the super version of the target of the model and said nothing about the source -- this is also a subject under consideration and we expect a lot of non-trivial links to \cite{superAKSZ}. 
And last but not least, we have studied ``independent'' observables to understand the phenomena related to them, but the physically relevant information is actually obtained when they are ``tested'' on other theories. A near-at-hand example would be supersymmetric Yang--Mills theory, by analogy with \cite{ACM15}. Those also merit a separate paper to figure out the details, and we expect interesting connection to supergravity. \\

\noindent 
\textbf{Acknowledgements.} \\
We thank Alexei Kotov, Ambar Sengupta, Anton Alekseev, Valentin Ovsienko, Camille Laurent-Gengoux and Anton Galaev for useful discussions and comments at various stages of this work. Some parts of this work have been presented at the Maxim Grigoriev's working group seminar at ITMP MSU. 
V.S. thanks Ambar Sengupta and Department of Mathematics of the University of Connecticut for hospitality in the beginning of this project, and appreciates O.C.'s personal grant that made this stay possible. Some of the meetings within this project were supported by the CNRS 80Prime GraNum project and its follow-ups, as well as the PHC Procope GraNum 2.0. with the University of Göttingen. 
We also thank Elena and Sergey Skoblia without whom this paper would take much longer to be finished. 

\newpage

\appendix

\section{Geometry of coadjoint orbits}  \label{sec:orbits}

\textbf{Coadjoint orbits.} 
Coadjoint orbits are the orbits of coadjoint action, they have the form
$ 
\OO_\lambda = \{Ad^*_g\lambda, \, g\in G\}.
$
We assume that $G$ is a matrix group and use matrix notation:
 $$
 Ad^*_g: \lambda \mapsto g^{-1}\lambda g,
 $$
where $\lambda \in \g^*$ and $g\in G$.
The stabilizer subgroups are  denoted by:
$$
H_\lambda = \{g\in G, \, Ad^*_g\lambda=g^{-1}\lambda g=\lambda \}.
$$
Then the coadjoint orbit is isomorphic to the homogeneous space $$
\OO_\lambda \cong G / H_\lambda,
$$
and it is convenient to introduce a canonical projection:
$$
\pi_\lambda: G \to \OO_\lambda.
$$
The group $G$ projects to every coadjoint orbit: we take a group element and map it to the image of $\lambda$ under the coadjoint action: $g\mapsto g^{-1}\lambda g = \mu$, where $\mu$ is  an arbitrary point on the orbit $\OO_\lambda$. 

The symplectic 2-form on $\OO_\lambda$ is constructed as follows. Consider a principal $H_\lambda$-bundle $G \to \OO_\lambda \cong G / H_\lambda$. First, for a given point on the orbit $\lambda \in \OO_\lambda$,  define a $\g$-valued 1-form on the group $G$: 
$\alpha = -Tr(\lambda g^{-1}dg) \in \Omega^1(G,\g)$. Here we also use matrix notation for Maurer-Cartan form $g^{-1}dg \in \Omega^1(G,\g)$. Then consider a 2-form on $G$ taking values in $\RR$: 
$$\omega_G= d\alpha = - Tr \lambda (g^{-1}dg)^2\in \Omega^2(G).$$
This 2-form is closed by definition and basic (i.e. horizontal and invariant under the right $G_\lambda$-action). Since $\Omega^\bullet(G)_{basic}\cong \Omega^\bullet(\OO_\lambda)$, it descends on the orbit $\OO_\lambda$ to the unique 2-form $\omega \in \Omega^2(\OO_\lambda)$. In more detail, there exists the unique $\omega \in \Omega^2(\OO_\lambda)$, such that $\pi^*_\lambda \omega = \omega_G \in \Omega^2(G)$. This 2-form on the orbit is $G$-invariant, non-degenerate and closed. It defines the symplectic structure on $\OO_\lambda$ and is also known as Kirillov--Kostant--Souriau form. 

Coadjoint orbits for which the KKS-form $\omega$ defines an integral cohomology class $[\omega]\in H^2(\mathcal{O}_\lambda,\mathbb{Z})$, are called integral. And Kirillov orbit method states that integral coadjoint orbits $\mathcal{O}_\lambda$ of a compact Lie group $G$ are in bijective correspondence with irreducible finite-dimensional representations of $G$: the orbit $\mathcal{O}_\lambda$ passing through the element $\lambda\in\mathfrak{h}^*$ corresponds to the representation $R_\lambda$ uniquely defined by the highest weight $\lambda\in\mathfrak{h}^*$.

\textbf{$\g^*$ is a Poisson manifold.}
Recall that the Lie algebra structure is defined on generators $T_a$ on $\g$ by the commutation relations $[T_a, T_b]=f_{ab}^cT_c$. The basis of $\g$ can be considered as coordinate functions on $\g^*$. 
On $\g^*$ there exists the unique Poisson bivector $\pi$ (\cite{KKS}), defining a map $\pi: T\g^* \to T\g$. In coordinates on $\g^*$, it has linear coefficients:
\begin{equation} \nonumber
    \pi = f_{ab}^cT_c \frac{\partial}{\partial T_a}\frac{\partial}{\partial T_b}.
\end{equation} 
The map $\pi: T\g^* \to T\g$ is not invertible in general. However, as a Poisson manifold, $(\g^*, \pi)$ can be represented as a disjoint union of smooth even-dimensional submanifolds: symplectic leaves. Constant values of Casimir functions are assigned to each leaf and thus parametrize the 
coadjoint orbits $\mathcal{O}_\lambda\subset \g^*$.

\section{Recap from super geometry}  \label{sec:superman}

The purpose of this Appendix is to give a fleeting overview of some folkloric facts from supergeometry, mostly to fix the notations and backup the constructions from the main part of this paper. More details can be found in any classical book on supergeometry, for example \cite{leites}.

\textbf{Superspaces and local models.} 

\noindent
As usual, a super vector space is decomposed into a direct sum 
$V = V_0 \oplus V_1$, coordinates on $V_i$ (i.e. elements of $V_i^*$) are of parity $i$.  This parity $p(\cdot) \in \mathbb Z_2$ enters the commutation relations, and defines the Koszul sign rule: $a \cdot b = \varepsilon(a, b)b \cdot a$, where 
 $\varepsilon(a, b) = (-1)^{p(a)p(b)}$.
One then constructs the tensor products and the  (super)symmetric algebra.
$$
T^k E = \underbrace{E \otimes \ldots \otimes E}_k \,.
$$
And for $\mathrm{T}(E) := \bigoplus_{k\ge 0} T^k$,  
$$
 S (E) := \mathrm{T}( E) / \left<  x\otimes y - \varepsilon(x, y) y \otimes x  \, | \, x, y\in E\right>\,.
$$

The functional space on $V$ is then $\mathcal{F}(V) = C^\infty(V_0) \otimes S(V_1)$, i.e. polynomials on $V_1$ with smooth coefficients on $V_0$. Such functional spaces supported over open sets covering a smooth manifold $M_0$ can be consistently glued together, giving rise to the following definition.

\noindent
\textbf{Def.}
A supermanifold $M^{k|l}$ is the data of $(M_0, \OO_M)$, where $M_0$ is a smooth manifold of dimension $k$ and $\OO_M$ is the structure sheaf, locally modelled as $\mathcal{F}(V)$
above, with $dim(V_0) = k$, $dim(V_1) = l$. \\

 Most of differential geometry works out of the box, up to the Koszul signs. One of the intricate points is related to measure and integration. For instance, not all the ingredients for the Stokes theorem  are automatically defined.   
 
 Out of all the rich data about supermanifolds, we will need the specific case of Lie groups and algebras (see e.g. 
\cite{Kac}), and for the structures related to the resulting gauge theories super symplectic and super Poisson structures are relevant, that we recall below.

\textbf{Supergroups and superalgebras}  

\noindent
\textbf{Def.}
A \emph{Lie superalgebra} is a vector superspace (over $\mathbb R$ or $\mathbb C$) 
$$
\g =\g_0 \oplus \g_1
$$
together with a super skew-symmetric bilinear operation of even parity
$$
[-,-]\colon \g\otimes \g\to \g\, \hspace{3mm}
[x,y]=(-1)^{p(x)p(y)+1}[y,x], \hspace{2mm} x,y\in\g,,
$$ 
which satisfies the super Jacobi identity
$$
\big[x, [y,z]\big]=\big[[x,y],z\big]+(-1)^{p(x)p(y)}
\big[y, [x,z]\big]
$$

\noindent
\textbf{Def.}
An \emph{invariant product} on Lie superalgebra $\g =\g_0 \oplus \g_1$ is a bilinear form $Tr(\cdot,\cdot)$ on $\g$, which is supersymmetric, non-degenerate, invariant, and can be even or odd. 

\noindent
A bilinear form on Lie superalgebra $\g$ is \\
-- \emph{supersymmetric} if 
$Tr(x, y) = (-1)^{p(x)p(y)}Tr(y, x)$, for all 
homogeneous $x$ and $y$.\\
-- \emph{non-degenerate} if $x \in \g$ satisfies $Tr(x, y) = 0$, for all $y\in \g$, then necessarily $x=0$.\\
-- \emph{invariant} if $Tr([x, y], z) = Tr(x, [y, z])$, for all $x, y, z \in \g$.\\
-- \emph{even} if $Tr(\g_0, \g_1) = Tr(\g_1, \g_0) = \{0\}$.\\
-- \emph{odd} if $Tr(\g_0, \g_0) = Tr(\g_1, \g_1) = \{0\}$.

\noindent
\textbf{Def.}
A \emph{Lie supergroup} $G$ is a group object in the category of supermanifolds. That is $G$ is a supermanifold on which the standard group operations (multiplication, inverse and identity) are defined as morphisms, and satisfy the standard group axioms. 

While the proper way to describe supergroups is to do it through the sheaf of functions to distinguish odd and even elements, a shorthand notation is sometimes: $G = G_{0}\times G_{1}$. As in the smooth case Lie supergroups and superalgebras are related by an integration / differentiation procedure.

\textbf{Super Poisson and super symplectic structures}  \\ 
\noindent
The standard construction of tensor fields from classical differential geometry can be extended to supermanifolds in a rather straightforward way. The key difference is obviously related to supersymmetry and superskewsymmetry: for each (skew)symmetric tensor field written in local coordinates the coefficients will be divided into two groups corresponding to odd and even variables. Not going into general details 
we mention only three important particular cases: \\
--- superskewsymmetric covariant 2-tensor field i.e. differential 2-form \\
--- supersymmetric covariant 2-tensor field i.e. metric \\
--- superskewsymmetric contravariant 2-tensor field i.e. bivector field. \\
All of them are allowed to be odd or even.

\noindent
\textbf{Def.} A \emph{supersymplectic form} is a closed and non degenerate differential 2-form on a vector superspace / supermanifold.

\noindent
\textbf{Def.} Given a supermanifold $M$, a \emph{super Poisson structure} on $\OO_M$ is a mapping 
$\{\cdot, \cdot\} \colon \OO_M \times \OO_M \to \OO_M$, such that $(\OO_M, \{,\})$ is a Lie superalgebra in the sense above, and any $\{f, \cdot\}$ acts as a superderivation. \\

\end{document}